\documentclass[manuscript]{aastex}
\usepackage{amsmath}

\shorttitle{Features of Tachyon-Dominated Cosmology}
\shortauthors{Martin and Redmount}

\begin{document}
\vskip -1.0in
\title{Observable Features of Tachyon-Dominated Cosmology}

\author{Audrey Claire Martin}
\affil{Northern Arizona University\\
	Department of Astronomy and Planetary Sciences\\
	South San Francisco Street\\
	Flagstaff, Arizona~~86011~~USA}
\email{acm586@nau.edu}


\author{Ian H.~Redmount}
\affil{Saint Louis University\\
	Department of Physics\\
	3511 Laclede Avenue\\
	Saint Louis, Missouri~~63103--2010~~USA}
\email{ian.redmount@slu.edu}

\begin{abstract}
A Friedmann-Robertson-Walker cosmological model dominated by
tachyonic---faster-than-light---dark matter can exhibit features similar
to those of a standard dark energy/dark matter or $\Lambda$CDM model.
It can undergo expansion which decelerates to a minimum rate, passes
through a ``cosmic jerk,'' then accelerates.  But some features of
a tachyon-dominated model are sufficiently distinct from those of
the standard model that the two possibilities might be distinguished
observationally.  As a demonstration of concept, the distance-redshift
relation of such a model is compared here with some observations of
Type~Ia supernovae.  Other measures of the \emph{third} time derivative of
the cosmic scale factor---the true cosmic jerk---might be found to test a
tachyonic-dark-matter hypothesis.
\end{abstract}


\section{Introduction}\label{sec01}

Over the past two decades it has become clear that the matter/energy
content of the cosmos must be dominated by constituents quite distinct
from the ordinary ``luminous'' matter of stars and planets.  These
are called ``dark matter'' if they obey equations of state similar to
those of familiar matter and energy, and ``dark energy'' if they do
not---the latter including cosmological-constant or vacuum-energy
contributions.  Despite considerable and ongoing efforts, however,
specific dark-matter and dark-energy components have not yet been
identified.  Hence, it remains possible to consider even exotic candidates.
For example, a gas of tachyons---faster-than-light particles, with spacelike
four-momenta---can drive an open (spatially hyperbolic) cosmolgical model
which expands from an initial singularity at a rate which decelerates to
a minimum value, passes through a ``cosmic jerk,'' then
accelerates~\citep{star2022}.  These are features that, in accord with
current observations, characterize what has become the ``standard''
cosmological-constant/cold-dark-matter or $\Lambda$CDM cosmological model.
The behavior of such a tachyon-dominated model differs sufficiently
from a $\Lambda$CDM model that the two might be distinguished observationally.
As a demonstration of concept, we present here some basic features of
such a tachyon-dominated model to be tested against observations.
A variety of more rigorous tests~\citep{kram2022, gopa2022} should be
possible.

A suitable tachyon-dominated model universe is described in Sec.~\ref{sec02}.
Numerical parameters sufficient to specify the model are obtained from a few
basic cosmological measurements in Sec.~\ref{sec03}.  The distance-redshift
relation for the model, which might be compared to a more extensive set of
observations, is shown in Sec.~\ref{sec04}.  A tachyon-dominated model
fitted to the same observational data, with its own predictions for the
basic cosmological measurements, is presented in Sec.~\ref{sec05}.
Conclusions drawn from these results are described in Sec.~\ref{sec06}.
\vfil\eject
\section{Cosmological Model}\label{sec02}

\begin{figure*}
\includegraphics{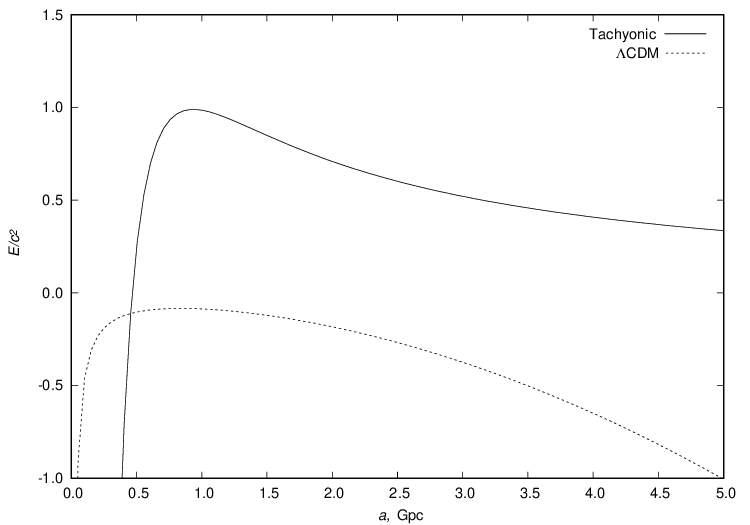}
\caption{\label{f01}Friedmann-Equation ``potential'' functions for
tachyon-dominated (solid curve) and dark-energy/dark-matter $\Lambda$CDM
(dashed curve) models.  Here ``energy''~$E$ refers to the terms of the
Friedmann Equation~\eqref{eq02}, in units of~$c^2$, and~$a$ is the scale
factor of the models, in gigaparsecs.}
\end{figure*}

The model in question is an open---spatially
hyperbolic---Friedmann-Robertson-Walker spacetime geometry.
It has line element
\begin{equation}
\label{eq01}
ds^2=-c^2\,dt^2+a^2(t)\left[d\chi^2+\sinh^2\chi\,\left(d\theta^2
+\sin^2\theta\,d\phi^2\right)\right]\ ,
\end{equation}
with comoving time coordinate~$t$, angular coordinates~$\chi$, $\theta$,
and~$\phi$, and scale factor or curvature radius~$a(t)$.  The Einstein
Field Equations applied to this metric imply the Friedmann Equation
\begin{equation}
\label{eq02}
\left(\frac{da}{dt}\right)^2-\frac{8\pi G}{3c^2}\,\rho\,a^2=+c^2\ ,
\end{equation}
with $\rho$ the energy density of the cosmic fluid and~$G$ Newton's constant.
The evolution of the model is determined by the form of the density~$\rho$
as a function of~$a$.  A thermal ensemble of free, noninteracting (i.e.,
dark-matter) tachyons gives rise to density~\citep{star2022}
\begin{equation}
\label{eq03}
\rho(a)=\frac{(\rho_0+{\cal M}_0c^2)a_0^4}{a^4}
-\frac{{\cal M}_0c^2a_0^3}{a^3}\ ,
\end{equation}
with $\rho_0$ the energy density and ${\cal M}_0$ an
invariant-mass\footnote{Tachyons, never at rest, are not properly
characterized by ``rest mass.''} density at some fiducial time~$t_0$---e.g.,
the present time---at which the scale factor has value~$a_0$.  The Friedmann
Equation~\eqref{eq02} takes the form
\begin{subequations}
\begin{equation}
\label{eq04a}
\left(\frac{da}{dt}\right)^2+\left(\frac{2B}{a}-\frac{A^2}{a^2}\right)
=+c^2\ ,
\end{equation}
\begin{equation}
\label{eq04b}
\hbox{with}\qquad A\equiv\left(
\frac{8\pi G\,(\rho_0+{\cal M}_0c^2)\,a_0^4}{3c^2}\right)^{1/2}
\end{equation}
\begin{equation}
\label{eq04c}
\hbox{and}\qquad B\equiv\frac{4\pi G{\cal M}_0\,a_0^3}{3}\ .
\end{equation}
\end{subequations}
The ``potential energy'' term on the left-hand side of Eq.~\eqref{eq04a}
is illustrated in Figure~\ref{f01}.  For parameter values giving $B<Ac$,
the model expands from an initial singularity ($a\to0^+$)
at a decreasing rate, passes through a minimum expansion rate at the peak
of the potential, then accelerates.  Unlike the standard $\Lambda$CDM model,
which ultimately continues to accelerate exponentially in time, this
tachyon-dominated model asymptotically approaches finite expansion rate
$da/dt=c$.  The tachyonic model illustrated in the figure has parameter
values given in Sec.~\ref{sec03}, following.  The $\Lambda$CDM model shown
has density
\begin{subequations}
\begin{equation}
\label{eq05a}
\rho_{\Lambda{\rm CDM}}(a)=\rho_\Lambda+\frac{\rho_{m0}a_0^3}{a^3}\ ,
\end{equation}
with constant vacuum-energy density
\begin{equation}
\label{eq05b}
\rho_\Lambda=6.06\times10^{-10}~\hbox{J/m}^3\ ,
\end{equation}
current-time matter density
\begin{equation}
\label{eq05c}
\rho_{m0}=2.60\times10^{-10}~\hbox{J/m}^3\ ,
\end{equation}
\end{subequations}
and the current-time scale factor from Eq.~\eqref{eq10e}, for purposes
of comparison.

The Friedmann Equation~\eqref{eq04a} can be solved in closed form.  The model
of interest here, satisfying $B<Ac$, is described via scale factor and time
\begin{subequations}
\begin{equation}
\label{eq06a}
a(\eta)=\frac{A}{c}\,\sinh\eta-\frac{B}{c^2}\,(\cosh\eta-1)
\end{equation}
and
\begin{equation}
\label{eq06b}
t(\eta)=\frac{A}{c^2}\,(\cosh\eta-1)-\frac{B}{c^3}\,(\sinh\eta-\eta)\ ,
\end{equation}
\end{subequations}
in terms of \textit{conformal time} parameter $\eta\in[0,+\infty)$.

\section{Model Parameters}\label{sec03}

This model is specified by three parameters, e.g., the parameter~$A$
setting the overall scale of the model, the dimensionless ratio
$\beta\equiv B/(Ac)$, and the current conformal-time value~$\eta_0$.
These are to be determined by fitting predictions of the model to a
suitable set of observations.  A simple example of such a procedure
is shown in Sec.~\ref{sec05}.  But as a starting point, simply to
test the feasibility of the model, the parameters can be set by matching
the values of three well-known cosmological quantities, so that other consequences
of the model can be compared to other data.  For example,
the current value of the Hubble parameter
\begin{subequations}
\begin{equation}
\label{eq07a}
\begin{aligned}[b]
H_0&\equiv\left.\frac{1}{a_0}\,\frac{da}{dt}\right|_{\eta=\eta_0}\\
&=\frac{c^2}{A}\,\frac{\cosh\eta_0-\beta\,\sinh\eta_0}
{[\sinh\eta_0-\beta\,(\cosh\eta_0-1)]^2}\ ,\\
\end{aligned}
\end{equation}
the current age of the universe
\begin{equation}
\label{eq07b}
t_0=\frac{A}{c^2}\,[(\cosh\eta_0-1)-\beta\,(\sinh\eta_0-\eta_0)]\ ,
\end{equation}
and the redshift~$z_j$ of the ``cosmic jerk,'' at which the acceleration
of cosmic expansion changes from negative to positive, given by
\begin{equation}
\label{eq07c}
\begin{aligned}[b]
1+z_j&\equiv\frac{a(\eta_0)}{A/(\beta c)}\\
&=\beta\,\sinh\eta_0-\beta^2\,(\cosh\eta_0-1)\\
\end{aligned}
\end{equation}
\end{subequations}
can be used.  Solving Eq.~\eqref{eq07c} for~$\beta$, then solving the combined
equation
\begin{equation}
\label{eq08}
H_0t_0=\left(\frac{\beta}{1+z_j}\right)^2\,(\cosh\eta_0-\beta\,\sinh\eta_0)
[(\cosh\eta_0-\beta\,\sinh\eta_0)+(\beta\eta_0-1)]
\end{equation}
for~$\eta_0$, then fixing~$A$ via Eq.~\eqref{eq07b} determines all three
model parameters.

Using the following sample values~\citep{hins2009,laha2010}:
\begin{subequations}
\begin{equation}
\label{eq09a}
\begin{aligned}[b]
H_0&=71.7~\frac{\hbox{km/s}}{\hbox{Mpc}}\\
&=2.32\times10^{-18}~\hbox{s}^{-1}\\
\end{aligned}
\end{equation}
and
\begin{equation}
\label{eq09b}
\begin{aligned}[b]
t_0&=1.37\times10^{10}~\hbox{yr}\\
&=4.32\times10^{17}~\hbox{s}\ ,\\
\end{aligned}
\end{equation}
and estimate
\begin{equation}
\label{eq09c}
z_j=0.500\ ,
\end{equation}
\end{subequations}
this procedure yields parameters
\begin{subequations}
\begin{equation}
\label{eq10a}
\beta=0.993\ ,
\end{equation}
\begin{equation}
\label{eq10b}
\eta_0=5.02\ ,
\end{equation}
and
\begin{equation}
\label{eq10c}
\begin{aligned}[b]
A&=(9.57\times10^{16}~\hbox{s})\,c^2\\
&=(0.930~\hbox{Gpc})\,c\ .\\
\end{aligned}
\end{equation}
These imply values
\begin{equation}
\label{eq10d}
\begin{aligned}[b]
B&=(9.50\times10^{16}~\hbox{s})\,c^3\\
&=(0.924~\hbox{Gpc})\,c^2\\
\end{aligned}
\end{equation}
and present-time scale factor
\begin{equation}
\label{eq10e}
a_0=1.40~\hbox{Gpc}
\end{equation}
\end{subequations}
for this version of the model.

\section{Distance-Redshift Relation}\label{sec04}

The distance-redshift relation is a feature of the model that might be
tested against astronomical observations.  Since incoming light rays in
a spacetime geometry with line element~\eqref{eq01} travel on trajectories
satisfying $d\chi=-d\eta$, the {\bf metric\/} distance~$\ell$ to a source
is given by
\begin{equation}
\label{eq11}
\ell=a_0(\eta_0-\eta_*)\ ,
\end{equation}
with $\eta_*$ the conformal-time parameter at emission of the signal, and
subscript zero denoting present-time values.  The redshift~$z$ of the source
is related to the scale factor~$a(\eta)$ thus:
\begin{equation}
\label{eq12}
1+z=\frac{a_0}{a(\eta_*)}\ .
\end{equation}
These can be combined with the model's scale factor~\eqref{eq06a} to obtain
distance-redshift relation
\begin{equation}
\label{eq13}
\begin{aligned}[b]
\ell(z)&=a_0\,\left[\eta_0-\alpha-\sinh^{-1}\left(
\frac{a_0\,\cosh\alpha}{(A/c)(1+z)}-\sinh\alpha\right)\right]\\
&=(1.40~\hbox{Gpc})\,\left\{2.19-\sinh^{-1}\left[8.53\left(
\frac{1.51}{1+z}-0.993\right)\right]\right\}\ .\\
\end{aligned}
\end{equation}
Here the parameter $\alpha\equiv\tanh^{-1}\beta$ is introduced.
Numerical values are taken from results~\eqref{eq10a}--\eqref{eq10e};
these imply value $\alpha=2.83$.

Metric distance~$\ell(z)$ is not measured directly.  Observations provide
{\it luminosity distances} \citep{mtw1973,lppt1975}, e.g.,
\begin{equation}
\label{eq14}
\begin{aligned}[b]
D_L(z)&\equiv\left(\dfrac{\cal L}{4\pi S}\right)^{1/2}\\
&=(1\times10^{-7}~\hbox{Gpc})\,\exp\left(\dfrac{\ln(10)\,(m-M)}
{5}\right)\\
&=a_0\,(1+z)\,\sinh[\ell(z)/a_0]\\
&=a_0\,(1+z)\,\sinh\left[\eta_0-\alpha-\sinh^{-1}\left(
\frac{a_0\,\cosh\alpha}{(A/c)(1+z)}-\sinh\alpha\right)\right]\\
&=(1.40~\hbox{Gpc})\,(1+z)\,\sinh\left\{2.19-\sinh^{-1}\left[8.53\left(
\frac{1.51}{1+z}-0.993\right)\right]\right\}\ ,\\
\end{aligned}
\end{equation}
for a source with intensity~$S$, luminosity~${\cal L}$, apparent magnitude~$m$,
and absolute magnitude~$M$.  The overall factor~$1+z$ incorporates redshift
and time-dilation effects, and the $\sinh$ function incorporates the curvature
correction appropriate to an open model with line element~\eqref{eq01}. 

\begin{figure*}
\includegraphics{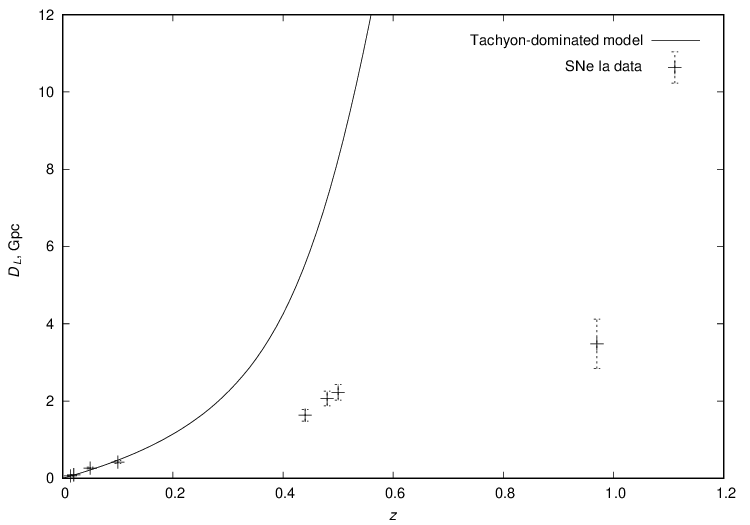}
\caption{\label{f02}Distance-redshift relation for the open
(spatially hyperbolic) tachyon-dominated model with parameters
from Eqs~\eqref{eq10a}--\eqref{eq10e}.  Data points are
adapted from~\citet{garn1998}, as displayed in Table~\ref{tbl01}}
\end{figure*}

While this relation may appear opaque, it is actually not so unreasonable.
It can be compared to measurements of Type~Ia supernovae, e.g.,
from~\citet{garn1998}.  Data adapted from Figure~3 of that work are shown
in Table~\ref{tbl01}, which displays redshift~$z$, distance moduli~$m-M$
\begin{table}[h]
\begin{center}
\caption{Redshift/Distance Data for Type~Ia Supernovae
from~\citet{garn1998}\label{tbl01}}
\begin{tabular}{cccc}
\\\tableline\tableline
$z$ & $m-M$ & $D_L$, Gpc & $\sigma_D$, Gpc \\
\tableline
0.014 & 34.00 & 0.0629 & 0.006 \\
0.020 & 34.80 & 0.0904 & 0.008 \\
0.050 & 37.20 & 0.2654 & 0.024 \\
0.100 & 38.30 & 0.4205 & 0.039 \\
0.440 & 41.83 & 1.6342 & 0.151 \\
0.480 & 42.40 & 2.0677 & 0.191 \\
0.500 & 42.59 & 2.2268 & 0.205 \\
0.970 & 44.15 & 3.4796 & 0.641 \\
\tableline
\end{tabular}
\end{center}
\end{table}
for apparent magnitudes~$m$ and absolute magnitudes~$M$,
distances~$D_L$ obtained via the second line of Eq.~\eqref{eq14},
and uncertainties~$\sigma_D$ estimated from the error bars shown
in~\citet{garn1998} and that equation. The comparison of model relation
and data is shown in Figure~\ref{f02}.
Here the model relation fits the low-redshift data points reasonably
well, but yields greater distances for the larger redshift values.
As a demonstration of concept, however, this shows that the tachyon-dominated
model could be tested agains such observations.  The parameters~$A$, $B$,
and~$\eta_0$---that is, the invariant mass, density, and temperature of the
tachyon gas, as well as the distance and time scales of the model---could be
fixed, as is done to evaluate the features of more familiar
models~\citep{hins2009}, by fitting the model to the data.

\section{Tachyon-Dominated Model Fitted to Data}\label{sec05}

\begin{figure*}
\includegraphics{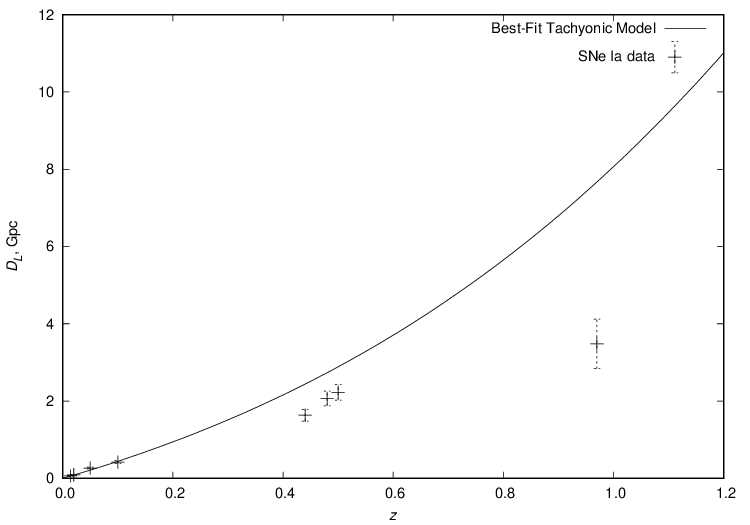}
\caption{\label{f03}Distance-redshift relation for the open tachyon-dominated
model with best-fit parameters given by Eqs.~\eqref{eq15a}--\eqref{eq15i}.
Data points are as displayed in Table~\ref{tbl01} and Fig.~\ref{f02}.}
\end{figure*}

Distance-redshift relation~\eqref{eq14} can be fitted to the data of
Table~\ref{tbl01} by adjusting the three model parameters~$a_0$, $\eta_0$,
and~$\alpha$.  The results of a Levenberg-Marquardt $\chi^2$-minimization
calculation~\citep{pftv1986a} are shown in Figure~\ref{f03}.
The parameters of the model shown are those obtained from a simple
Monte Carlo calculation:  Ten iterations of the fitting program are carried
out, with normally-distributed variations introduced into the
data.~\citep{pftv1986b}  The means and standard deviations of the resulting
parameter values~\citep{pftv1986c} yield
\begin{subequations}
\begin{equation}
\label{eq15a}
a_0=(2.97\pm0.71)~\hbox{Gpc}\ ,
\end{equation}
\begin{equation}
\label{eq15b}
\eta_0=3.33\pm0.41\ ,
\end{equation}
and
\begin{equation}
\label{eq15c}
\alpha=1.28\pm0.35\ ,
\end{equation}
and model features
\begin{equation}
\label{eq15d}
\beta=0.856\pm0.093\ ,
\end{equation}
\begin{equation}
\label{eq15e}
\begin{aligned}[b]
A&=[(1.07\pm0.26)\times10^{17}~\hbox{s}]\,c^2\\
&=[(1.04\pm0.25)~\hbox{Gpc}]\,c\ ,\\
\end{aligned}
\end{equation}
\begin{equation}
\label{eq15f}
\begin{aligned}[b]
B&=[(0.896\pm0.28)\times10^{17}~\hbox{s}]\,c^3\\
&=[(0.871\pm0.28)~\hbox{Gpc}]\,c^2\ ,\\
\end{aligned}
\end{equation}
present-time Hubble parameter
\begin{equation}
\label{eq15g}
H_0=(71.7\pm3.9)~\hbox{km/s/Mpc}\ ,
\end{equation}
current age
\begin{equation}
\label{eq15h}
\begin{aligned}[b]
t_0&=(1.31\pm0.15)\times10^{10}~\hbox{yr}\\
&=(4.13\pm0.46)\times10^{17}~\hbox{s}\ ,\\
\end{aligned}
\end{equation}
and cosmic-jerk redshift
\begin{equation}
\label{eq15i}
z_j=1.62\pm1.40\ .
\end{equation}
\end{subequations}

The fit shown in Figure~\ref{f03} has $\chi^2=100.$, or $\chi^2$ per degree of
freedom~$20.0$; the iterations of the Monte Carlo calculation have
average~$\chi^2$ value~144., or $\chi^2$ per degree of freedom~28.8.
These suggest, as the graph does, at best a modest fit to the data.  It may be
that the small errors of the low-$z$ distance values force a close fit to these
points, leaving larger discrepancies at the higher~$z$ values.  Clearly a larger,
more carefully curated data set is needed for a more thoroughgoing test of the
tachyonic model; one such study is in progress.~\citep{kram2022}

But results~\eqref{eq15g} and~\eqref{eq15h}, in particular, show that the
tachyon-dominated model can yield cosmological features close, but not
identical, to those obtained from more standard models---such as
Eqs.~\eqref{eq09a} and~\eqref{eq09b}.  Hence, these results serve as a
``demonstration of concept,'' indicating that the tachyonic dark-matter
model might ultimately serve as a novel alternative to the now standard
$\Lambda$CDM cosmology.

\section{Conclusions}\label{sec06}

Despite their somewhat fanciful nature, tachyons might actually be a viable
dark-matter candidate.  An open Friedmann-Robertson-Walker spacetime with
mass/energy content dominated by a tachyon gas can show the same general
features---expansion from an initial singularity decelerating to a minimum
rate, then subsequently accelerating---as a standard $\Lambda$CDM~model.
More specific features of such a tachyon-dominated model can be compared
with observations---sufficient perhaps to justify more careful and
thorough investigation.

One important distinction between the tachyonic and $\Lambda$CDM models is
that the latter continues expanding exponentially in time, while the former
asymptotically approaches constant expansion rate $da/dt=c$.  Hence,
determination of not just the first and second time derivatives of the
scale factor~$a$, but also its third derivative---the true ``cosmic
jerk''---might be crucial to distinguishing between the two possibilities.
Furthermore, the standard model must asymptotically approach the geometry of
de~Sitter~\citep{desi1917a,desi1917b} spacetime, while the tachyonic model
must approach the Milne~\citep{miln1932} universe, which is \emph{flat
spacetime.} This might imply that the quantum-state spaces for
elementary-particle fields in the two models are different~\citep{redm1999},
hence, that the two possible cosmologies might even be distinguished by close
examination of physics at the very smallest scales.

\clearpage

\end{document}